\documentstyle[aps,preprint]{revtex}
\begin{document}
\draft
\title{Elusive Order Parameters for Non-Abelian Gauge Theories}
\author{Hoi-Kwong Lo}
\address{
 School of Natural Sciences, Institute for Advanced Study, Olden Lane,\\
 Princeton, NJ 08540, U.S.A.
}
\date{\today}
\preprint{IASSNS-HEP-95/4}
\maketitle
\mediumtext
\begin{abstract}
We construct a set of order parameters for non-Abelian
gauge theories which probe directly the unbroken group and are
free of the deficiencies caused by quantum fluctuations and
gauge fixing which have plagued all previous attempts.
These operators can be used to map out the phase diagram of a non-Abelian
gauge theory.
\end{abstract}
\narrowtext
A convenient way to investigate the phase diagram of a gauge theory
is to construct a set of order parameters which exhibit non-analytical
behavior at phase boundaries. For
{\em Abelian} gauge theories,
Preskill and Krauss \cite{PreKra} have
made a successful construction by invoking
the Aharonov-Bohm effect \cite{AhaBoh}
between cosmic strings and charged particles.\footnote{Another
promising order
parameter--the vacuum overlap order parameter--has been proposed by
Fredenhagen and Marcu\cite{FreMar}. See\cite{qft,Lo} for discussions.}
Generalization to non-Abelian
gauge theories turned out to be very elusive. In spite of
much progress in our understanding of the subtler aspects of
non-Abelian gauge theories
\cite{qft,Bavade,AlfMar}, it proves difficult to formulate
a general procedure that unambiguously specifies the unbroken group.
The key difficulty is quantum fluctuations.
We need a framework that takes full account of the effects of
virtual cosmic string loops (magnetic flux tubes).
Any useful Aharonov-Bohm experiment necessarily proceeds in two stages:
calibration and measurement. World sheets of virtual string loops
can wrap around test charges, thus changing their states relative
to other charges in the universe. Consequently, repeated flux measurements
with test charges will not necessarily agree.
It is also important for us to gauge fix. An element of a unbroken
group has no invariant meaning unless the value of the Higgs
field is specified.
We address these important issues in this Letter and
construct a set of order parameters that will probe directly and
unambiguously the unbroken group of a non-Abelian gauge theory.
We borrow our ideas from
Alford {\em et al.} \cite{qft}
who developed but immediately rejected a set
of operators which were plagued by quantum fluctutations.

We emphasize that the idea of using the Aharonov-Bohm
effect to probe the unbroken group is rather
general. Moreover, the associated subtleties due
to gauge fixing and quantum fluctuations of
virtual magnetic flux tubes addressed in this Letter are
general consequences of a partial symmetry breakdown of a non-Abelian
gauge theory and the non-Abelian Aharonov-Bohm effect.
These issues arise no matter the subgroup is discrete or
continuous.
However, for simplicity, we will discuss
the following scenario only.
Consider a simply-connected gauge group $K$ which
is spontaneously broken into a discrete non-Abelian group $G$.
This symmetry breaking leads to the existence of stable vortices
(in 2+1 dimensions) and cosmic strings (in 3+1 dimensions)
labelled by the elements of $G$. (Topologically stable vortices
are classified by homotopy group $\pi_1(K/G)$ \cite{Vilenk,Preski}.
It follows
from the exact homotopy sequence that $\pi_1(K/G)
\simeq G$ for a simply connected $K$. To be more precise,
the spectrum of
stable vortices only spans $G$. An element of $G$ may be unstable
to decay into two or more vortices with the same total flux.
This is, however, of no interest to our discussion.)
Depending on its Higgs structure and the parameters of the Higgs potential,
the symmetry group $G$ may be further broken into a subgroup $H$.
We would like to construct a set of order parameters to test if
such a symmetry breakdown has occurred and if it does, what is
the unbroken group $H$?

In a free $G$ charge phase,
charged particles generally scatter
off cosmic strings non-trivially.
When a charged particle in the initial state
$| u \rangle$ (and representation $\nu$)
winds around a string loop of flux $a$, its state becomes $ D^{\nu} (a)
| u \rangle$. This non-Abelian Aharonov-Bohm effect can be invoked
to probe the phase diagram of a non-Abelian gauge theory. More
concretely,
we proceed as follows:
prepare (or calibrate) a set of classical vortices,
one for each element of $G$.
Measure the
non-Abelian Aharonov-Bohm phases acquired by charged particles
which wind around the various vortices that we have prepared.
Read off the spectrum of stable vortices from the results of
our experiments and decide if symmetry breaking has occurred.

In the free charge phase, each element of $G$ is associated with
a stable pointlike vortex. Indeed, a test charge in the
representation $\nu$
which traverses a
$g$-string acquires a phase $D^{\nu}(g)$ as expected.
We, therefore, conclude that
the gauge group $G$ is unbroken.
On the other hand, if $G$ is broken into a subgroup $H$,
the elements of $G$ that are not in $H$ are not associated with
isolated cosmic strings, but with strings that are boundaries of
domain walls. Such domain walls are unstable and will decay via
spontaneous nucleation of string loops \cite{qft,Vilenk,Preski}.
Consider the insertion of a classical string worldsheet of flux $a \notin H$.
Holes eventually appear in the wall bounded to the $a$ string and
collide with one another. Ultimately, the one $b$
with the least string tension will dominate the decay and a charged
particle scattering off the composite string will therefore measure
a flux $ab^{-1} \in H$ rather than $a$. Consequently,
if we measure the Aharonov-Bohm
phases acquired by test particles which traverse the various calibrated
vortices, we find that the fluxes of the vortices are always elements
of $H$ and conclude that a symmetry breaking from $G$ to $H$
has occurred.

It was suggested in \cite{KraWil} that when a $U(1)$ gauge symmetry
is spontaneously broken into $Z_N$, the discrete $Z_N$ charge
$Q_{\Sigma}^*$ contained in a closed surface $\Sigma^*$ can
still be measured via the Gauss law:
\begin{equation}
F(\Sigma^*)= \exp \left({2 \pi i \over N} Q_{\Sigma^*} \right)
=\exp  \left( {2 \pi i \over
Ne} \int_{\Sigma^*} E \cdot ds \right). \label{gauss}
\end{equation}
($F(\Sigma^*)$ is closely related to the 't Hooft loop operator \cite{tHooft}.)
$F(\Sigma^*)$ inserts a classical cosmic string source on the world
sheet $\Sigma^*$.

Now we turn to the operator which introduces classical charges into the
system. An obvious choice would be the Wilson loop
operator $W^{\nu}(C)$ where $\nu$ is an irreducible representation of
the gauge group, $G$. The Aharonov-Bohm
phase acquired by a charged particle which scatters off a cosmic string
is represented by $F(\Sigma^*) W^{\nu}(C)$ where $C$ and $\Sigma^*$ have
a non-trivial linking number.
One might naively expect that
$F(\Sigma^*) W^{\nu}(C)$ is the order parameter.
This is not quite correct
because quantum mechanical fluctuations near
the surface
$\Sigma^* $ cause an area law decay of the {\em modulus\/} of
$F(\Sigma^*) \sim \exp \left(- \kappa A({\Sigma^*}) \right)$. Fortunately,
the {\em phase\/} of $F(\Sigma^*)$ remains unscreened and we can isolate
it by dividing out its vacuum expectation value and obtain
$F(\Sigma^*) \over \langle F(\Sigma^*) \rangle $\cite{PreKra} .
Similarly, quantum fluctuations also
lead to the exponential decay of the expectation value of $W(C)$.
Therefore, the true order parameter for Abelian gauge theories is
\cite{PreKra}
\begin{equation}
A^{\nu}(\Sigma^{*}, C) = {F(\Sigma^{*})
W^{\nu}(C) \over  \langle F(\Sigma^*) \rangle
\langle W^{\nu}(C)\rangle}.
  \label{abelian}
\end{equation}
In the free $Z_N$ charge phase, the order parameter (for the fundamental
representation) gives
\begin{equation}
\lim \langle A(\Sigma^{*}, C) \rangle
 = \exp \left({2 \pi i \over N} k( \Sigma^* ,C) \right) .
\label{free}
\end{equation}
Here the limit is taken with $\Sigma^* $ and $C$ increasing to infinite
size, and with the closest approach of $\Sigma^*$ to $C$ also approaching
infinity; $ k( \Sigma^* ,C)$ denotes the linking number of the surface
$\Sigma^*$ and the loop $C$. (Note that other than these
requirements, the value of $A$ is independent of the details of
$\Sigma^*$ and $C$. Thus, we can probe the unbroken group by
performing a finite number of thought experiments.)
On the other hand, if there are no free
$Z_N$ charges, then we have
\begin{equation}
\lim \langle A(\Sigma^{*}, C) \rangle
 = 1 .
\label{nofree}
\end{equation}
The non-analytical behavior of $ A(\Sigma^{*}, C)$ guarantees that the
two phases are separated by a well-defined phase boundary.
To probe the realisation of
any Abelian discrete gauge symmetry, we just
consider the operators $F_a(\Sigma^*)$
for each element $a \in G$.

When the gauge group is non-Abelian, the flux, $h$ of a string
has no gauge invariant meaning. One can imagine
choosing an arbitrary base point $x_0$, setting up a basis of test particles
and calibrating the flux of a string by scattering these
particles of known transformation properties from it along a particular
path.
Another important issue is gauge fixing.
Suppose we are interested in studying the symmetry breaking of
$G$ into $H$. After symmetry breaking, strings
with fluxes
$h$ and $ghg^{-1}$ ($g \in G$) are typically {\em not} gauge equivalent
to each other.
To test whether symmetry breaking has occurred, one has to
choose a field $\phi$ as a candidate for the Higgs field,
gauge fix $\phi = \phi_0$ at $x_0$ and consider $H( \phi_0)$ and
conjugacy classes and representations of $H$.

In the lattice formulation, it is convenient to put a string world sheet
on a closed
surface $\Sigma^* $ on the dual lattice. Let $\Sigma$ be the set of
plaquettes threaded by $\Sigma^*$.
Now, for each plaquette $P$
in $\Sigma$, we choose a path, $l_P$, that runs from the base point $x_0$ to a
corner of the
plaquette \cite{qft}. Calibration of the plaquette is done
along the path $l_P P l_P^{-1} $.
More concretely, suppose that the plaquette action is
\begin{equation}
   S^{(R)}_{gauge, P}= - \beta \chi^{(R)} (U_P) + c.c. \label{oldgauge}
\end{equation}
   where $R$ is some representation of the gauge group that defines the theory.
The insertion of $F_a(\Sigma^{*}, x_0, \{l_P\},\phi_0)$ modifies the action
on each plaquette in $\Sigma$ to
\begin{equation}
 S^{(R)}_{gauge, P} \to
 - \beta \chi^{(R)} (V_{l_P}a V_{l_P}^{-1} U_P) + c.c. \label{newgauge}
\end{equation}
where
\begin{equation}
V_{l_P} = \prod_{l \in l_P} U_l.  \label{path}
\end{equation}

Now we turn to the operator which introduces classical charges into the
system. Having gauge
fixed the Higgs candidate
$\phi= \phi_0$
at $x_0$, all information of non-Abelian Aharonov-Bohm effect is encoded
in
the {\em untraced\/} Wilson loop operator
\begin{equation}
U^{(\nu)}(C, x_0,\phi_0)= D^{(\nu)} \left( \prod_{l \in C} U_l \right)
\label{untrace}
\end{equation}
 where $C$ is a closed loop based at $x_0$, $\nu$
is an irreducible representation of the gauge group $G$.
The matrix elements of
$U^{(\nu)}(C, x_0,\phi_0)$ can, in principle, be determined by interfering
charged particles in the
representation $\nu$
that traverse $C$ with those that stay at the base point \cite{Bais,AlCoMa}.
Just like
$F_a$, the operator $U^{(\nu)}(C, x_0, \phi_0)$ is not
gauge invariant.

If we did not gauge fix $\phi= \phi_0$ at the base point,
global gauge tranformations by the group $G$ would be allowed.
Thus, by the Schur's lemma,
$\langle U^{\nu}(C, x_0)
\rangle = \lambda I$. Notice that an irreducible representation
of $G$ is typically reducible in $H$. The result $\langle U^{\nu}(C, x_0)
\rangle = \lambda I$ means that it would not
be possible to resolve the various irreducible representations
of $H$ in the
decomposition of an irreducible
representation of $G$. This is clearly wrong. We conclude that it
is crucial to perform gauge fixing.

Returning to the operator $F$, so far we have been vague about the choice
of $\{l_P\}$. As it turns out, this is of greatest importance.
It was noted in Ref. \cite{qft} that in a phase with
free $G$ charges, and
in the leading order of
weak coupling perturbation theory,
the operator
\begin{eqnarray}
\langle A^{\nu}_{a}
( \Sigma^{*} ,x_0,\{l_P\};C) \rangle
&=&{ \langle F_a(\Sigma^{*}, x_0, \{l_P\}) U^{(\nu)}(C, x_0) \rangle \over
\langle F_a(\Sigma^{*}, x_0, \{l_P\}) \rangle
\langle tr U^{(\nu)}(C, x_0)}\rangle \nonumber \\
&=& {1 \over n_{\nu}} D^{\nu} \left( a^{k (\Sigma^{*}, C)} \right)
\label{order}
\end{eqnarray}
 where ${k (\Sigma^{*}, C)}$ is the linking number
of the surface $\Sigma^{*}$ and the loop $C$ and the limit that
$\Sigma^{*}$ and $C$ are infinitely large and far away is taken.
(Note that
Alford {\em et al.} overlooked the importance of gauge fixing
in their definitions of $F_a$ and $U^{\nu}$.)
However, owing to quantum fluctuations,
higher order terms in the weak coupling expansion may spoil
this result \cite{qft}.
The dominant contribution in a weak coupling expansion
comes from configurations with a low density of frustrated
plaquettes (i.e., a low density of virtual string loops).
Alford {\em et al.} implicitly chose
the long tails, $\{l_P\}$, from the
plaquettes of $\Sigma$ in such a way that all of
them finally merge together at some point $y_0$ which is
far away from the base point and is not on the Wilson loop.
Unfortunately, this choice is vulnerable to quantum fluctuations.
Consider in (2+1) dimensions
a virtual vortex-antivortex pair whose worldline is non-trivially linked to
the union of three objects:
the Wilson loop, the tails and the string loop under calibration. (FIG. 1.)
This will conjugate the measured flux relative the calibrated value.
In the weak coupling expansion, such a configuration
has a single excited link on the path that connects $x_0$ to $\Sigma^*$.
This causes (in three spacetime
dimensions) the excitation of
four plaquettes and is suppressed by terms that are {\em independent\/} of the
size of $\Sigma^*$ and $C$ or the separation between them.
Thus,
higher order corrections render the flux uncertain
up to conjugation and
this operator is useless
as
an order parameter. This was the conclusion drawn by Alford {\em et al.}
\cite{qft} .

Such a conclusion is unwarranted
as it is based on their implicit choice of $\{ l_P \}$.
Any useful Aharonov-Bohm
experiment to determine the flux of a string loop
necessarily proceeds in two stages: calibration (with the operator
$F(\Sigma^*)$)
and subsequent measurement (with the operator $W(C)$) .
Both stages involve interference experiments with two beams
of charged particles
one of which traverses the string loop while the other just
sits at the base point.
To construct an order parameter for non-Abelian gauge theories,
the effects of virtual
string loops need to be considered. In the choice made by Alford
{\em et al.},
the particles used for measurement have their
worldlines along the Wilson loop $C$ whereas the particles for
calibration are kept in another box whose worldline runs along the
long chain from $x_0$ and $y_0$ that are common to all the tails,
but distinct from the Wilson loop.
In other words, the particles for calibration and those for measurement
are stored in separate boxes. It is only because of the
decoupling of the two processes that quantum fluctuations can spoil
the results. The key issue is that repeated
Aharonov-Bohm experiments do not nessarily agree.
A virtual vortex-antivortex pair may
spontaneously nucleate, wrap around the box which contains the calibrating
particles and annihilate, thus changing the state of the calibrating
particles relative the ones
used for subsequent
measurement. Since the calibrating and measuring particles
are in different states, the outcomes of the
Aharonov-Bohm experiments done with
these two different types of
particles will generally be different.

To be more precise, a particle
initially in a pure state $|u \rangle$ becomes
a mixed state
\begin{equation}
\rho= {1 \over n_{[b']}}\sum_{b \in [b']} D^{\nu}(b)|u \rangle \langle
u| D^{\nu} (b^{-1}) ,
\label{mixed}
\end{equation}
after traversing a charge-zero string loop in the conjugacy class
$[b']$.
If this particle then travels around an $a$-vortex, the result
of the interference experiment
appears to show that the flux of the vortex is an incoherent
superposition of $b^{-1}a b$ for $b \in [b']$ \cite{qft}.
In conclusion,
test particles may fail to recover the calibrated flux $a$ of
the classical string that the operator $F_a(\Sigma^*)$ introduces.

The resolution is simple. Keep the particles for both
calibration and subsequent measurement in the {\em same} box.
In the definition of $F$,
calibration is done for all plaquettes threaded by the
string loop. Physically, this means that we
continuously calibrate the string loop by sending particles
around it. For the subsequent measurement, one beam of particles
should be kept in the box while the other beam winds around the
calibrated string. (The two beams are wave packets.)
The lattice realization of our choice of $\{ l_P\}$ is shown in FIG. 2.
Now a large portion of the Wilson loop
close to $x_0$ represents the worldline of
the box containing all test particles.
Since continuous calibration of the flux of string is done by
sending particles around the string loop,
the tails (calibration paths)
are chosen in such a way that many of them beginning from
the base point
are initially on the Wilson loop and branch out one by one
from it.
In Ref. \cite{Lo}, we argue rigorously that this construction
overcomes all difficulties caused by quantum fluctuations.

It is crucial not
to send out the two beams of
particles for subsequent measurement too closely spaced in time.
Otherwise, virtual string loops may wrap around the two beams,
thus changing the particles for measurement without affecting those
for calibration.
Virtual string loops can, of course, wind around the box
containing all the particles, but this will affect both the calibration
and measurement processes and lead to no net change.
It is also possible for virtual string loops to
wind around one of the two beams, say, that for the measurement. However,
this incoherent effect will go away on average
if we are willing to repeat
independent identical experiments many times.

As remarked before, it is of utmost importance to gauge
fix $\phi= \phi_0$ at the base point in the definitions of $W$ and $F$.
With our new choice of $\{ l_P \}$ and gauge fixing,
detailed arguments in Ref. \cite{Lo} verify the non-analytical
behavior of the operator
\begin{equation}
\langle A^{\mu}_{a}
( \Sigma^{*}, x_0,\phi_0, \{l_P\};C) \rangle
 ={ \left\langle {1 \over |H|} \sum_{h \in H(\phi_0)}
F_{hah^{-1}}(\Sigma^{*}, x_0, \{l_P\},\phi_0 ) tr U^{\mu}(C, x_0, \phi_0)
\right\rangle
\over \left\langle {1 \over |H|} \sum_{h \in H(\phi_0)}
F_{hah^{-1}}(\Sigma^{*}, x_0, \{l_P\},\phi_0 ) \right\rangle
\left\langle
tr U^{\mu}(C,x_0, \phi_0) \right\rangle} .
\label{correctop}
\end{equation}
Note that we sum over only the elements
and an irreducible representation $\mu$ of $H$. By dealing
with the subtleties due to quantum fluctuations and
gauge fixing squarely , we see clearly \cite{Lo} how a gauge group $G$ is
reduced to an effective symmetry group $H$ at low energies. Subsequent
symmetry breaking of $H$ can be studied in a similar manner.

In conclusion, we have constructed a set of order parameters for non-Abelian
gauge theories.
The study of the symmetry breakdown of
$S_3$ into $Z_2$ is sketched in Ref. \cite{Lo}, which also discusses
the {\em coherent} insertion of two or more string loops and contains
derivations of many results stated in this Letter.

We are indebted to J. Preskill for bringing the issue
of quantum fluctuations to our attention
and for stimulating discussions.
Numerous helpful conversations with S.
Adler, M. Bucher,
H. F. Chau, R. Dick,
K.-M. Lee, J. March-Russell, P. McGraw, F. Wilczek,
M. de Wild Propitius and P. Yi are also gratefully acknowledged.
This work was supported in part by DOE DE-FG02-90ER40542.

\begin{figure}
\caption{Suppose all $l_P$ merge together at some point $y_0$ {\em not}
on the Wilson loop
before reaching the base point.
A worldline of virtual vortex conjugates the calibrated flux.}
\label{fig1}
\end{figure}
\begin{figure}
\caption{Lots of long tails initially lie on the Wilson loop $C$.
They eventually
branch out from it one by one and never intersect one another afterwards.
Moreover, the Wilson loop never comes close to retracing itself.
To conjugate the calibrated flux without affecting the measurement,
a virtual string loop must wrap around each tail after the branching
out of each tail
from the Wilson loop. Such a configuration becomes energetically
costly as $C$ and $\Sigma^*$ become large.}
\label{fig2}
\end{figure}
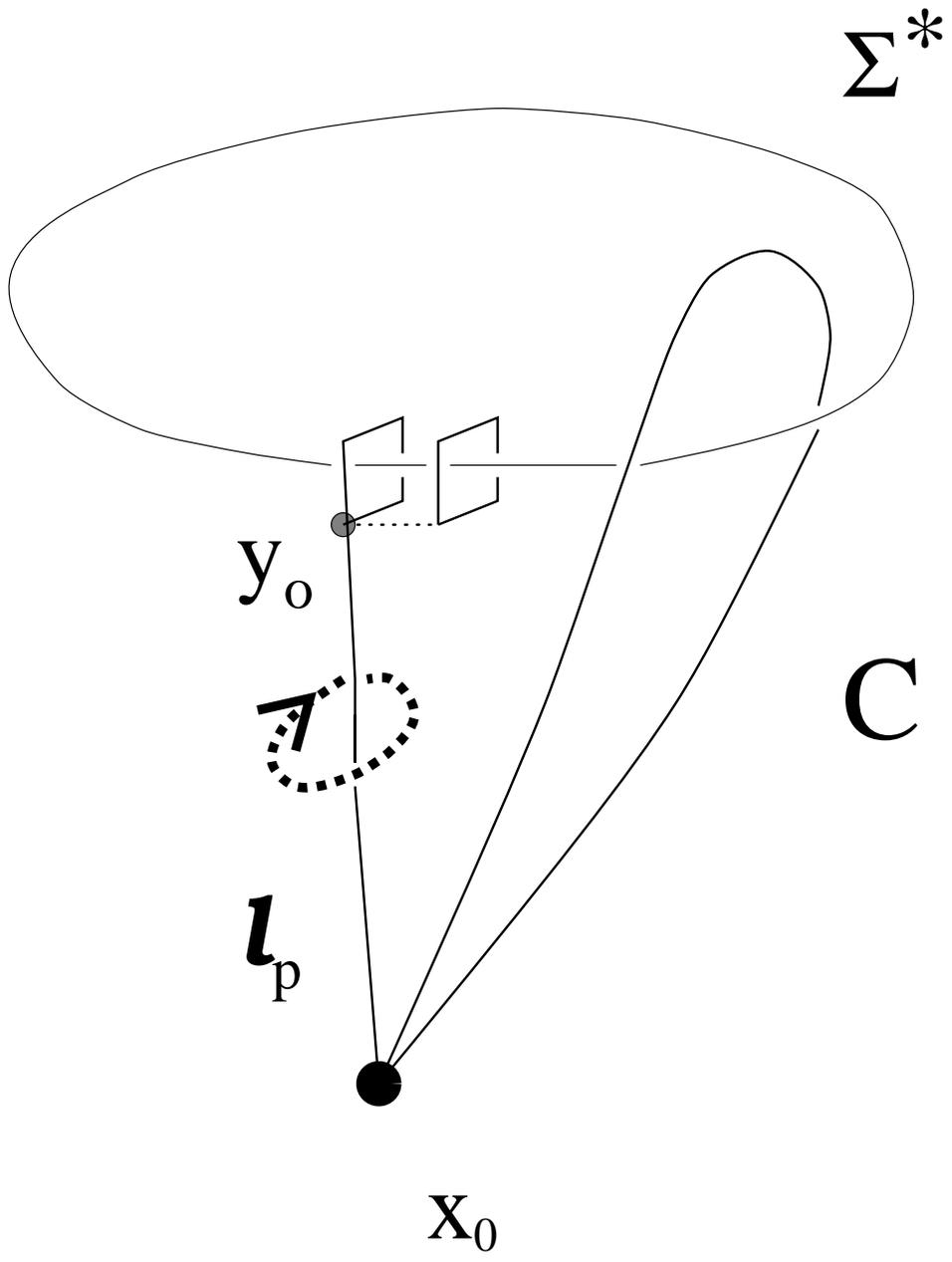
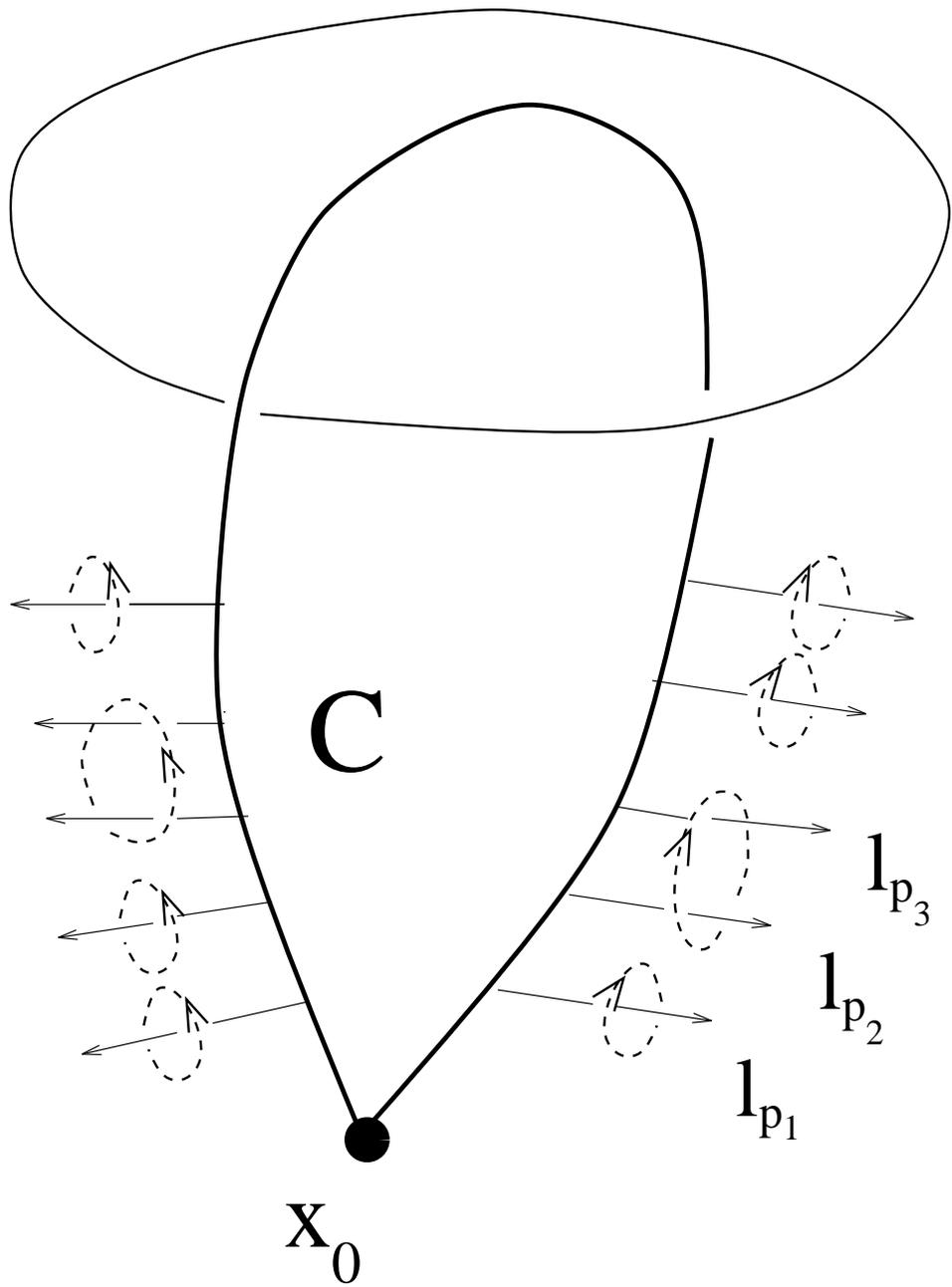
\end{document}